\documentclass[12pt]{iopart}
\usepackage{iopams}
\usepackage{graphicx}
\usepackage{graphics}

\begin{document}

\newcommand{\vc}[1]{\mathbf{#1}}

\title{Phonon dispersion and anomalies in one-layer high-temperature superconductors}
\author{Thomas Bauer, Claus Falter}
\ead{falter@uni-muenster.de}
\address{Institut f\"ur Festk\"orpertheorie, Westf\"alische Wilhelms-Universit\"at,\\
Wilhelm-Klemm-Str.~10, 48149 M\"unster, Germany}
\date{\today}

\begin{abstract}
Phonon dynamics, charge response and the phonon density of states are
calculated for the high-temperature superconductors HgBa$_{2}$CuO$_{4}$
and Bi$_{2}$Sr$_{2}$CuO$_{6}$ within a microscopic model for the
electronic density response. The results are compared with previous
calculations for La$_{2}$CuO$_{4}$ and Nd$_{2}$CuO$_{4}$. Our main
focus is on the phononanomalies which are connected with the
high-frequency oxygen bond-stretching  modes (OBSM) found before in our
calculations for $p$-doped La$_{2}$CuO$_{4}$ and $n$-doped
Nd$_{2}$CuO$_{4}$. We investigate the question if the characteristic
softening of the OBSM and the related strong coupling to the electrons
is also present in HgBa$_{2}$CuO$_{4}$ and Bi$_{2}$Sr$_{2}$CuO$_{6}$.
In particular the importance of the contribution of the more
delocalized Cu$4s$ state besides the localized Cu$3d$ state on the
softening is investigated and the different anticrossing behaviour due
to the presence of several phonon modes with the same symmetry as the
OBSM is studied. This makes the identification of the anomalies quite
complicate. Furthermore, the influence of electronic polarization
processes at ions out of the CuO plane in the ionic layers on the
phonon dynamics is calculated. In this context the qualitative type of
charge response, i.e. the presence or not of possible metallic charge
fluctuations at the Hg or Bi ion, respectively,  linked via the apex
oxygen to the CuO plane, proves to be very sensitive for certain phonon
modes. All the calculations are compared to the experimental results
available so far. The latter, however, are rather incomplete for the Hg
and the Bi compound.
\end{abstract}


\maketitle

\section{Introduction}\label{SecOne}
In recent literature there is increasing evidence that the
electron-phonon interaction (EPI) is significant in the cuprate based
high-temperature superconductors (HTSC's) and it is argued that phonons
might play an important role for the electron dynamics in the HTSC's;
see, e.g. Refs.
\cite{Lanzara01,Cuk05,Pint05,Gweon04,Zhou05,Reznik06,Lee06,Zhao07}. It
is known from the experiments for some time that the frequencies of the
oxygen bond-stretching modes (OBSM) are strongly renormalized
(softened) upon doping in the $p$-doped cuprates
\cite{Pint05,Pint98,Pint99,Reich96,Uchi04,Fuku05,Pint06,McQ01}. Quite
recently first evidence of an anomalous dispersion of the OBSM also has
been reported in a Bi cuprate \cite{Graf08} which, among other things,
will be investigated theoretically in this work. Anomalous softening of
the OBSM, not existing in the undoped insulating state, also has been
observed in the $n$-doped metallic state of Nd$_{2}$CuO$_{4}$
\cite{d'Astu02,d'Astu03,Braden05}. The phonon dispersion and the charge
response in Nd$_{2}$CuO$_{4}$ have been calculated in \cite{Bauer08}
and the result compares well with the measured anomalous dispersion.

All these findings support a generic nature of the OBSM phonon
anomalies in the HTSC's. The frequency renormalization and a
corresponding increase of the linewidths point to a strong coupling of
these phonons to the charge carriers. Deviations from the presumably
generic dispersion for the OBSM in form of a sharp softening have
recently been reported in inelastic neutron scattering (INS) and high
resolution inelastic x-ray scattering (IXS) experiments, respectively,
for La$_{2-x}$Ba$_{x}$CuO$_{4}$ at $x = 0.125$ \cite{Reznik06} and $x =
0.14$ \cite{d'Astu08}. This has been related to the possibility of
stripe formation in this compound.

According to our calculations the origin of the phonon anomalies is a
strong {\it nonlocal} EPI due to specific screening effects in form of
a polarization created by metallic charge fluctuations (CF's) on the
outer shells of the Cu and O$_{xy}$ ions in the CuO plane; see, e.g.
\cite{Falter05,Falter93,Falter06}. This type of screening
characteristic for a strongly inhomogeneous electronic system like the
cuprates with a strong component of ionic binding and a corresponding
density distribution which is mainly concentrated at the ions in the
CuO plane leads for the $p$-doped cuprates \cite{Falter05,Falter06} and
also for the $n$-doped Nd$_{2}$CuO$_{4}$ \cite{Bauer08} to an anomalous
behaviour of the dispersion.

The specific form of screening yields for the OBSM in metallic
La$_{2}$CuO$_{4}$ and Nd$_{2}$CuO$_{4}$ a {\it downward} dispersion of
the anomalous longitudinal branch with high frequencies along the
(1,0,0) and the (1,1,0) direction, respectively. In case of
Nd$_{2}$CuO$_{4}$, however, the dispersion of the branches with high
frequencies comprising the anomalous OBSM is more complex than in
La$_{2}$CuO$_{4}$ because there are in contrast to La$_{2}$CuO$_{4}$
several branches of the same symmetry type nearby in energy leading to
complex anticrossing phenomena \cite{Bauer08}. Such a possibility is
also investigated for HgBa$_{2}$CuO$_{4}$ and Bi$_{2}$Sr$_{2}$CuO$_{6}$
in this work.

The anomalous dispersion cannot be understood within a typical lattice
dynamical model, like the shell model, extended by a homogeneous
electron gas screening that is {\it diagonal} in reciprocal space. This
type of dielectric response always yields an {\it increasing}
dispersion when passing from the center of the Brillouin zone (BZ) into
the zone, because of an incomplete screening of the phonon-induced
changes of the Coulomb potential at shorter distances. In contrast, the
anomalous softening of the OBSM in particular of the oxygen
half-breathing mode ($\Delta_1/2$ anomaly halfway the $\Delta \sim
(1,0,0)$ direction) and the full oxygen breathing mode (O$_{\rm B}^{X}$
at the $X$ point of the BZ) is a result of an "overscreening" of the
phonon induced changes of the Coulomb potential. The overscreening
effect has been shown by our calculations to be due to nonlocally
excited CF's localized at the Cu and O$_{xy}$ sites in the CuO plane
accompanied by (dynamic) charge ordering in form of localized stripes
of alternating sign in the plane \cite{Falter05,Falter97}. Expressed in
terms of the density response matrix in reciprocal space such a strong
nonlocal EPI effect and "overscreening" of the OBSM is related to the
{\it off-diagonal elements} (local field effect) being apparently very
important in the cuprates as revealed by the anomalous softening found
in our studies.

In our calculations a sufficiently broad set of orbital degrees of
freedom (Cu$3d$, Cu$4s$, O$2p$), the full three dimensional long range
Coulomb interaction as well as the short range local repulsion of the
electrons, especially the important on-site repulsion mediated by the
localized Cu$3d$ orbital, is considered quantitatively. All the
resulting couplings arising in the dynamical matrix are microscopically
well defined and can be calculated.

From our investigations, see, e.g. Ref. \cite{Bauer09a}, we extract a
strong EPI for the phonon anomalies $\Delta_1/2$ and O$_{\rm B}^{X}$,
which can well be treated in {\it adiabatic} approximation. Moreover,
we find a very strong nonadiabatic enhancement of the coupling compared
to the adiabatic case for phonon-like modes, like O$^{Z}_{z}$ (axial
oxygen breathing mode at the $Z$ point of the BZ) , from a small {\it
nonadiabatic} region around the $c$-axis. In this region poor dynamical
screening due to the slow charge dynamics around the $c$-axis of the
long range polar Coulomb interaction of the ions leads in metallic
LaCuO$_{4}$ to low-energy plasmons mixing with certain interlayer
phonons of the same symmetry. For a discussion of phonon-plasmon mixing
in the cuprates, see also \cite{Alex92}. An essential nonlocal
interaction of holes with $c$-axis polarized optical phonons in the
HTSC's has also been reported in \cite{Alex96,Hardy09}.

Our calculations show that besides the more delocalized Cu$4s$ state
the short range part of the Coulomb interaction $U_{d}$ related to the
localized Cu$3d$ orbital is particular important for the $\Delta_1/2$
and O$_{\rm B}^{X}$ mode polarized and propagating in the CuO plane
while the strong long range nonlocal polar Coulomb interaction plays an
important role for O$^{Z}_{z}$ and similar $\Lambda_{1}$ phonons
polarized and propagating along the $c$-axis. Thus, the relevance, of
both, short range electron repulsion $U_{d}$ in the correlated Cu$3d$
state as well as long range polar coupling of the electrons to the
phonons, is demonstrated by our calculations of the phonon anomalies in
the cuprates. Consequently, any quantitative study of the physics of
the normal  or superconducting state of the cuprates should consider
electron correlations and strong nonlocal EPI on an equal footing.

The paper is organized as follows. In section 2 the theory and modeling
is shortly reviewed. Section 3 presents the calculations for the phonon
dynamics, charge response and the phonon density of states of
HgBa$_{2}$CuO$_{4}$ and Bi$_{2}$Sr$_{2}$CuO$_{6}$ and a comparison with
earlier calculations of LaCuO$_{4}$ and Nd$_{2}$CuO$_{4}$. The main
focus is on the OBSM phonon anomalies in the various materials and the
effect of the charge response in the ionic layers on the dispersion. A
summary of the paper is given in section 4 and the conclusions are
drawn.

\section{Theory and modeling}\label{SecTwo}
In the following a brief survey of the theory and modeling is
presented. A detailed description can be found in \cite{Falter93} and
in particular in \cite{Falter99} where the calculation of the coupling
parameters of the theory is presented.

The local part of the electronic charge response and the EPI is
approximated in the spirit of the {\it quasi-ion approach}
\cite{Bauer08,Falter88} by an ab initio rigid ion model (RIM) taking
into account covalent ion softening in terms of (static) effective
ionic charges calculated from a tight-binding analysis. The
tight-binding analysis supplies these charges as extracted from the
orbital occupation numbers $Q_{\mu}$ of the $\mu$ (tight-binding)
orbital in question:
\begin{equation}\label{Eq1}
Q_{\mu} = \frac{2}{N} \sum\limits_{n\vc{k}} |C_{\mu n} (\vc{k})|^2.
\end{equation}
$C_{\mu n} (\vc{k})$ stands for the $\mu$-component of the eigenvector
of band $n$ at the wavevector $\vc{k}$ in the first BZ; the summation
in equation \eref{Eq1} runs over all occupied states and $N$ gives the
number of the elementary cells in the (periodic) crystal.

In addition, scaling of the short range part of certain pair potentials
between the ions is performed to simulate further covalence effects in
the calculation in such a way that the energy-minimized structure is as
close as possible to the experimental one \cite{Falter95}. Structure
optimization and energy minimization is very important for a reliable
calculation of the phonon dynamics through the dynamical matrix. Taking
just the experimental structure data as is done in many cases in the
literature may lead to uncontrolled errors in the phonon calculations.

The RIM with the corrections just mentioned then serves as an unbiased
reference system for the description of the HTSC's and can be
considered as a first approximation for the insulating state of these
compounds because of the strong ionic nature of bonding in the
cuprates. Starting with such an unprejudiced rigid reference system
non-rigid electronic polarization processes are introduced in form of
more or less localized electronic charge-fluctuations (CF's) at the
outer shells of the ions. Especially in the metallic state of the
HTSC's the latter dominate the {\em nonlocal} contribution of the
electronic density response and the EPI and are particularly important
in the CuO planes. In addition, \textit{anisotropic}
dipole-fluctuations (DF's) are admitted in our approach
\cite{Falter99,Falter02}. They prove to be specifically of interest for
the ions in the ionic layers mediating the dielectric coupling and for
the polar modes. Thus, the basic variable of our model is the ionic
density which is given in the perturbed state by
\begin{equation}\label{Eq2}
\rho_\alpha(\vc{r},Q_\lambda, \vc{p}_\alpha) =
\rho_\alpha^0(r) + \sum_{\lambda}Q_\lambda \rho_\lambda^{\rm CF}(r)
+ \vc{p}_\alpha \cdot
\hat{\vc{r}} \rho_\alpha^{\rm D}(r).
\end{equation}
$\rho_\alpha^0$ is the density of the unperturbed ion, as used in the
RIM, localized at the sublattice $\alpha$ of the crystal and moving
rigidly with the latter under displacement. The $Q_\lambda$ and
$\rho^{\rm CF}_\lambda$ describe the amplitudes and the form-factors of
the CF's and the last term in equation \eref{Eq2} represents the
dipolar deformation of an ion $\alpha$ with amplitude (dipole moment)
$\vc{p}_\alpha$ and a radial density distribution $\rho_\alpha^{\rm
D}$. $\hat{\vc{r}}$ denotes the unit vector in the direction of
$\vc{r}$. The $\rho^{\rm CF}_\lambda$ are approximated by a spherical
average of the orbital densities of the ionic shells calculated in LDA
taking self-interaction effects (SIC) into account. The dipole density
$\rho_\alpha^{\rm D}$ is obtained from a modified Sternheimer method in
the framework of LDA-SIC \cite{Falter99}. All SIC-calculations are
performed for the average spherical shell in the orbital-averaged form
according to Perdew and Zunger \cite{Perdew81}. For the correlation
part of the energy per electron the parametrization given in
\cite{Perdew81} has been used.

The total energy of the crystal is obtained by assuming that the
density can be approximated by a superposition of overlapping densities
$\rho_\alpha$. The $\rho_\alpha^0$ in equation \eref{Eq2} are also
calculated within LDA-SIC taking environment effects, via a Watson
sphere potential and the calculated static effective charges of the
ions into account. The Watson sphere method is only used for the oxygen
ions and the depth of the Watson sphere potential is set as the
Madelung potential at the corresponding site. Such an approximation
holds well in the HTSC's \cite{Falter95,Krak98}. Finally, applying the
pair-potential approximation we get for the total energy:
\begin{equation}\label{Eq3}
E(R,\zeta) = \sum_{\vc{a},\alpha} E_\alpha^\vc{a}(\zeta)
+\frac{1}{2}\sum_{(\vc{a},\alpha)\neq(\vc{b},\beta)}
\Phi_{\alpha\beta}
\left(\vc{R}^\vc{b}_\beta-\vc{R}^\vc{a}_\alpha,\zeta\right).
\end{equation}
The energy $E$ depends on both the configuration of the ions $\{R\}$
and the electronic (charge) degrees of freedom (EDF) $\{\zeta\}$ of the
charge density, i.e. $\{Q_\lambda\}$ and $\{\vc{p}_\alpha\}$ in
equation \eref{Eq2}. $E_\alpha^\vc{a}$ are the energies of the single
ions. $\vc{a}$, $\vc{b}$ denote the elementary cells and $\alpha$,
$\beta$ the corresponding sublattices. The second term in equation
\eref{Eq3} is the interaction energy of the system, expressed in terms
of \textit{anisotropic} pair-interactions $\Phi_{\alpha\beta}$. Both
$E_\alpha^\vc{a}$ and $\Phi_{\alpha\beta}$ in general depend upon
$\zeta$ via $\rho_\alpha$ in equation \eref{Eq2}.

The pair potentials in equation \eref{Eq3} can be seperated into long
range Coulomb contributions and short range terms, for details see e.g.
Ref \cite{Falter99}.

From the adiabatic condition
\begin{equation}\label{Eqnew4}
\frac{\partial E(R,\zeta)}{\partial \zeta} = 0
\end{equation}
the electronic degrees of freedom $\zeta$ can be eliminated, an
expression for the atomic force constants can be given and from this
the dynamical matrix in harmonic approximation can be derived as
\begin{eqnarray}\label{Eq4}\nonumber
t_{ij}^{\alpha\beta}(\vc{q}) &=
\left[t_{ij}^{\alpha\beta}(\vc{q})\right]_{\rm RIM}\\ &-
\frac{1}{\sqrt{M_\alpha M_\beta}} \sum\limits_{\kappa,\kappa'}
\left[B^{\kappa\alpha}_i(\vc{q}) \right]^{*} \left[C^{-1}(\vc{q})
\right]_{\kappa\kappa'} B^{\kappa'\beta}_j(\vc{q}).
\end{eqnarray}
The first term on the right hand side denotes the contribution from the
RIM. $M_\alpha$, $M_\beta$ are the masses of the ions and $\vc{q}$ is a
wave vector from the first BZ.

The quantities $\vc{B}(\vc{q})$ and $C(\vc{q})$ in equation \eref{Eq4}
represent the Fourier transforms of the electronic coupling
coefficients and are calculated from the energy in equation \eref{Eq3},
i.e.
\begin{eqnarray}\label{Eq5}
\vc{B}_{\kappa\beta}^{\vc{a}\vc{b}} &= \frac{\partial^2
E(R,\zeta)}{\partial \zeta_\kappa^\vc{a} \partial R_\beta^\vc{b}},
\\\label{Eq6} C_{\kappa\kappa'}^{\vc{a}\vc{b}} &= \frac{\partial^2
E(R,\zeta)}{\partial \zeta_\kappa^\vc{a} \partial
\zeta_{\kappa'}^\vc{b}}.
\end{eqnarray}
$\kappa$ denotes the EDF (CF and DF in the present model, see equation
\eref{Eq2}) in an elementary cell. The $\vc{B}$ coefficients describe
the coupling between the EDF and the displaced ions (bare
electron-phonon coupling), and the coefficients $C$ determine the
interaction between the EDF. The phonon frequencies
$\omega_\sigma(\vc{q})$ and the corresponding eigenvectors
$\vc{e}^\alpha(\vc{q}\sigma)$ of the modes $(\vc{q}\sigma)$ are
obtained from the secular equation for the dynamical matrix in equation
\eref{Eq4}, i.e.
\begin{equation}\label{Eq7}
\sum_{\beta,j} t_{ij}^{\alpha\beta}(\vc{q})e_j^\beta(\vc{q}) =
\omega^2(\vc{q}) e_i^\alpha(\vc{q}).
\end{equation}
The Eqs. \eref{Eq4}-\eref{Eq7} are generally valid and, in particular,
are independent of the specific model for the decomposition of the
perturbed density in equation \eref{Eq2} and the pair approximation in
equation \eref{Eq3} for the energy. The lenghty details of the
calculation of the coupling coefficients $\vc{B}$ and $C$ cannot be
reviewed in this paper. They are presented in \cite{Falter99}. In this
context we note that the coupling matrix $C_{\kappa\kappa'}(\vc{q})$ of
the EDF-EDF interaction, whose inverse appears in equation \eref{Eq4}
for the dynamical matrix, can be written in matrix notation as
\begin{equation}\label{Eq8}
C = \Pi^{-1} + \widetilde{V}.
\end{equation}
$\Pi^{-1}$ is the inverse of the {\em proper polarization part} of the
density response matrix and contains the kinetic part to the
interaction $C$ while $\widetilde{V}$ embodies the Hartree and
exchange-correlation contribution, because they are related to the
second functional derivatives with respect to the density $\rho$ of the
kinetic energy and the exchange-correlation energy, respectively
\cite{Falter93}. $C^{-1}$ needed for the dynamical matrix and the EPI
is closely related to the linear density response matrix and to the
inverse dielectric matrix $\varepsilon^{-1}$, respectively.

Only very few attempts have been made to calculate the phonon
dispersion and the EPI of the HTSC's using the linear response method
in form of density functional perturbation theory (DFPT) within LDA
\cite{Sav96,Wang99,Bohnen03,Giustino08}. These calculations correspond
to calculating $\Pi$ and $\widetilde{V}$ in DFT-LDA and for the
\textit{metallic} state only. On the other hand, in our microscopic
modeling DFT-LDA-SIC calculations are performed for the various
densities in equation \eref{Eq2} in order to obtain the coupling
coefficients $\vc{B}$ and $\widetilde{V}$. Including SIC is
particularly important for localized orbitals like Cu$3d$ in the
HTSC's. Our theoretical results for the phonon dispersion
\cite{Falter05,Falter06,Bauer09a,Falter02}, which compare well with the
experiments, demonstrate that the approximative calculation of the
coupling coefficients in our approach is sufficient, even for the
localized Cu$3d$ states. Written in matrix notation we get for the
density response matrix the relation
\begin{equation}\label{Eq9}
C^{-1} = \Pi(1+\widetilde{V}\Pi)^{-1} \equiv \Pi \varepsilon^{-1},
\hspace{.7cm} \varepsilon = 1 + \widetilde{V}\Pi.
\end{equation}
The CF-CF submatrix of the matrix $\Pi$ can approximately be calculated
for the metallic (but not for the undoped and underdoped) state of the
HTSC's from a TBA of a single particle electronic bandstructure. In
this case the electronic polarizability $\Pi$ in tight-binding
representation reads:
\begin{eqnarray}\nonumber
\Pi_{\kappa\kappa'}&(\vc{q},\omega=0) = -\frac{2}{N}\sum\limits_{n,n',\vc{k}}
\frac{f_{n'}(\vc{k}+\vc{q})
-f_{n}(\vc{k})}{E_{n'}(\vc{k}+\vc{q})-E_{n}(\vc{k})}
\times \\\label{Eq10} &\times \left[C_{\kappa n}^{*}(\vc{k})C_{\kappa
n'}(\vc{k}+\vc{q}) \right] \left[C_{\kappa' n}^{*}(\vc{k})C_{\kappa'
n'}(\vc{k}+\vc{q}) \right]^{*}.
\end{eqnarray}
$f$, $E$ and $C$ in equation \eref{Eq10} are the occupation numbers,
the single-particle energies and the expansion coefficientes of the
Bloch-functions in terms of tight-binding functions.

The self-consistent change of an EDF at an ion induced by a phonon mode
$(\vc{q} \sigma)$ with frequency $\omega_\sigma(\vc{q})$ and
eigenvector $\vc{e}^\alpha(\vc{q}\sigma)$ can be derived in the form
\begin{eqnarray}\label{Eq11}
\delta\zeta_\kappa^\vc{a}(\vc{q}\sigma) & = & \left[-\sum_\alpha
\vc{X}^{\kappa\alpha}(\vc{q})\vc{u}_\alpha(\vc{q}\sigma)\right]
e^{i\vc{q}\vc{R}_\kappa^\vc{a}}\nonumber\\
& \equiv & \delta\zeta_\kappa(\vc{q}\sigma)e^{i\vc{q}\vc{R}^\vc{a}},
\end{eqnarray}
with the displacement of the ions
\begin{eqnarray}\label{Eq12}
\vc{u}_\alpha^{\vc{a}}(\vc{q}\sigma) & = &
\left(\frac{\hbar}{2M_\alpha\omega_\sigma(\vc{q})}
\right)^{1/2}\vc{e}^\alpha(\vc{q}\sigma)e^{i\vc{q}\vc{R}^\vc{a}}\nonumber\\
& \equiv & \vc{u}_\alpha(\vc{q}\sigma)e^{i\vc{q}\vc{R}^\vc{a}}.
\end{eqnarray}
The self-consistent response per unit displacement of the EDF in
equation \eref{Eq11} is calculated in linear response theory as:
\begin{equation}\label{Eq13}
\vc{X}(\vc{q}) = \Pi(\vc{q})\varepsilon^{-1}(\vc{q})\vc{B}(\vc{q}) =
C^{-1}(\vc{q})\vc{B}(\vc{q}).
\end{equation}
The generalization for the quantity $\Pi$ in Eqs. \eref{Eq8} and
\eref{Eq9} needed for the kinetic part of the charge response in the
nonadiabatic regime, where dynamical screening effects must be
considered, can be achieved by adding $(\hbar\omega+i \eta)$ to the
differences of the single-particle energies in the denominator of the
expression for $\Pi$ in equation \eref{Eq10}. Other possible
nonadiabatic contributions to $C$ related to dynamical
exchange-correlation effects and the phonons themselves are beyond the
scope of the present approach. Using equation \eref{Eq9} for the
dielectric matrix, $\varepsilon$, and the frequency-dependent version
of the irreducible polarization part, $\Pi$, according to equation
\eref{Eq10}, the free-plasmon dispersion is obtained from the
condition,
\begin{equation} \label{Eq14}
{\rm det} [\varepsilon_{\kappa\kappa'} (\vc{q}, \omega)] = 0 .
\end{equation}
The coupled-mode frequencies of the phonons and the plasmons must be
determined self-consistently from the secular equation \eref{Eq7} for
the dynamical matrix which now contains the frequency $\omega$
implicitly via $\Pi$ in the response function $C^{-1}$. Such a
nonadiabatic approach is necessary for a description of the interlayer
phonons and the charge-response within a small region around the
$c$-axis as performed in \cite{Falter05,Bauer09a}.

\section{Results and discussion} \label{SecThree}
\subsection{Ionic reference system and structural data}
For a definitive discussion of the renormalization of the phonon modes
introduced by the nonlocal EPI effects related to the electronic
polarization processes of CF and DF type, a quantitative reference
model for the calculation of the phonon dispersion based essentially on
the important component of ionic binding is necessary. A suitable model
sketched in section 2 which represents approximately the local EPI
effects is provided by the \textit{ab initio} rigid-ion model, extended
by covalent ion softening and scaling of the short range part of
certain pair potentials.

\begin{table}[b]
\centering
\begin{tabular}{cccc}
La$_2$CuO$_4$ & Nd$_2$CuO$_4$ & HgBa$_{2}$CuO$_{4}$ &
Bi$_{2}$Sr$_{2}$CuO$_{6}$\\\hline
 La: 2.28+       & Nd: 2.35+        & Hg: 1.44+      & Bi: 2.28+\\
              &        & Ba: 1.69+      & Sr: 1.85+\\
 Cu: 1.22+    & Cu: 1.22+   & Cu: 1.70+      & Cu: 1.22+\\
 O$_{xy}$: 1.42-  & O$_{xy}$: 1.42-   & O$_{xy}$: 1.56-& O$_{xy}$: 1.42-\\
 O$_{z}$: 1.47-   & O$_{z}$: 1.54-                 & O$_{z}$: 1.70- & O$_{z}$: 1.47-\\
                 &                  &                & O3: 1.85-
\end{tabular}
\caption{Static effective charges of the ions.}\label{tab01}
\end{table}

\begin{table}
\centering
\begin{tabular}{rlrlrlrl}
\multicolumn{2}{c}{La$_2$CuO$_4$ \cite{Longo73}} &
\multicolumn{2}{c}{Nd$_2$CuO$_4$ \cite{Mueller75}} &
\multicolumn{2}{c}{ HgBa$_{2}$CuO$_{4}$}\cite{Wagner93} & \multicolumn{2}{c}{Bi$_{2}$Sr$_{2}$CuO$_{6}$}\cite{Torardi88}\\
\hline
$a=$ & 3.763       & $a=$ & 4.155     & $a=$ & 3.8995    & $a=$ & 3.741  \\
 & (3.810)  &   & (3.945)    & &(3.875)   & & (3.798) \\
$c=$ & 13.197    & $c=$ &12.102   & $c=$ &9.521     & $c=$& 23.490 \\
 & (13.240)    & &(12.171)  & &(9.513)    &  &(24.662) \\
$c/a=$ & 3.5    & $c/a=$ & 2.9   & $c/a=$ & 2.4     & $c/a=$& 6.3 \\
 & (3.5)    & &(2.9)  & &(2.5)    &  &(6.5) \\
$\tau({\rm O}_z)=$& 0.185  & $\tau({\rm O}_z)=$& 0.250  & $\tau({\rm O}_z)=$& 0.313  & $\tau({\rm O}_z)=$& 0.109  \\
& (0.182) &  & (0.250) &  & (0.294) &  & (0.105) \\
$\tau({\rm La})=$& 0.134   & $\tau({\rm Nd})=$& 0.153   & $\tau({\rm Ba})=$ &0.183  & $\tau({\rm O3})=$& 0.199  \\
 & (0.138)  &  & (0.149)  &  & (0.198)  &  & (0.186) \\
& & & & & & $\tau({\rm Bi})=$ & 0.189  \\
& & &  & & & & (0.184) \\
& & & & & & $\tau({\rm Sr})=$ &0.068  \\
& & &  & & & & (0.071) \\
Cu-O$_{xy}$=& 1.881 & Cu-O$_{xy}$=& 2.078 & Cu-O$_{xy}$=& 1.950 & Cu-O$_{xy}$=& 1.870 \\
 &(1.895) & &(1.972) & &(1.940) & & (1.896)\\
 Cu-O$_{z}$=& 2.441 & Cu-O$_{z}$=& 3.670 & Cu-O$_{z}$=& 2.980 & Cu-O$_{z}$=& 2.556 \\
 &(2.407) & &(3.622) & &(2.793) & & (2.585)\\
 O$_z$-La=& 2.382 & O$_z$-Nd=& 2.389 & O$_z$-Hg=& 1.800 & O$_z$-Bi=& 1.893 \\
 &(2.381) & &(2.323) & &(1.966) & & (1.944)\\
\end{tabular}
\caption{Structural data for the investigated compounds. For each
lattice parameter, the first row shows the calculated data and the
second row the experimental data in brackets. The lattice constants $a$
and $c$ and the distances between the ions are given in units of \AA,
the relative translation $\tau$ is in units of $c$. In Nd$_2$CuO$_4$,
$\tau({\rm O}_z)$ is fixed by symmetry. The experimental references are
listed following the name of the compounds. }\label{tab02}
\end{table}

A modified ionic description is quite general a good starting point for
the physics in the cuprates in particular for their insulating state.
In this context we also refer to the work in
\cite{Catlow98,Islam88,Zhang91}. In our calculations we found a
suitable set of static effective charges given in \tref{tab01} which
together with covalent scaling leads to a good overall agreement with
the experimental structural data, shown in \tref{tab02}. The charges of
the Cu and the O$_{xy}$ ions in the CuO plane are taken identical for
La$_{2}$CuO$_{4}$, Nd$_{2}$CuO$_{4}$ and Bi$_{2}$Sr$_{2}$CuO$_{6}$.
They result from a tight-binding analysis of the \textit{ab initio}
band structure in La$_{2}$CuO$_{4}$ \cite{Falter95}. This means that
the CuO plane is treated as generic for these materials as far as the
effective charges are concerned. On the other hand, for
HgBa$_{2}$CuO$_{4}$ these charges have to be chosen more ionic,
otherwise the phonon modes are too low in frequency as compared with
the experiment. In all the materials covalent scaling of the short
range pair potential between the Cu and the O$_{xy}$ ions in the plane
is very essential and in case of HgBa$_{2}$CuO$_{4}$ and
Bi$_{2}$Sr$_{2}$CuO$_{6}$ in addition between the apex oxygen O$_{z}$
and Hg and O$_{z}$ and Bi, respectively.

This is consistent with a significant shorter distance of the
O$_{z}$-Hg and O$_{z}$-Bi link as compared to O$_{z}$-La or O$_{z}$-Nd
in La$_{2}$CuO$_{4}$ and Nd$_{2}$CuO$_{4}$, respectively (O$_{z}$-Hg:
1.97 \AA; O$_{z}$-Bi: 1.94 \AA; O$_{z}$-La: 2.38 \AA; O$_{z}$-Nd: 2.32
\AA). Thus, the overlap of the orbital densities in the O$_{z}$-Hg and
O$_{z}$-Bi bond is sufficiently large and consequently metallic CF's
between the partially occupied O$_{z}$2p and Hg$6s$ orbitals and the
O$_{z}$2p and Bi$6p$ orbitals, respectively, can be expected. The
important effect of such CF's on certain phonon modes will be discussed
in section 3.2.

\subsection{Phonon dispersion, charge response and phonon density of states}
The calculation of the phonon dispersion in this section is based on
the dynamical matrix given in equation \eref{Eq4} within adiabatic
approximation which has been shown to be a good approximation for all
phonon modes outside a small region around the $c$-axis, where, on the
other hand, a nonadiabatic approach is necessary
\cite{Falter05,Bauer09a}. As discussed, e.g. in Ref. \cite{Bauer09a},
the nonadiabatic region around the $c$-axis is so small that it cannot
be resolved by INS experiments or other probes so far. Thus, only an
average of the dispersion can be measured along the $\Lambda \sim
(0,0,1)$ direction for modes of $\Lambda_{1}$ symmetry because of their
coupling to a low lying $c$-axis plasmon as discussed in
\cite{Bauer09a}. Consequently, all results for the
$\Lambda_{1}$-phonons propagating and polarized along the $c$-axis
calculated in adiabatic approximation or reported in the experiments
must be interpreted as an average over the small region with a
nonadiabatic charge repsonse.

Anisotropic DF's being particular important along the $c$-axis and for
ions in the ionic layers \cite{Bauer08,Falter02,Falter03} have been
taken into account in all computations of the phonon modes in this
paper.

The subsequent calculations are representative for the well doped
metallic state of the materials where a fully developed Fermi surface
exists. Investigations of the charge response and phonon dynamics of
the underdoped pseudogap state and the insulating state of the cuprates
have been performed in \cite{Bauer08,Falter06,Falter02}. These
calculations are based on a microscopic modeling of the electronic
polarizability in terms of rigorous orbital specific
compressibility-incompressibility sum rules for
$\Pi_{\kappa\kappa'}(\vc{q})$ in the long wave-length limit in order to
discriminate between the charge response of the different electronic
sates of the cuprates.

In the following we discuss the effect of the renormalization of the
phonon dispersion by metallic CF's which are of particular importance
for the anomalous high-frequency oxygen bond-stretching  modes OBSM. In
our earlier calculations of the phonon anomalies and the phonon
dispersion of La$_{2}$CuO$_{4}$ and Nd$_{2}$CuO$_{4}$
\cite{Bauer08,Falter06,Falter02,Falter00} we have applied successfully
a parameterized description ($\Pi$-model) for the electronic
polarizabality matrix $\Pi_{\kappa\kappa'}$ in equation \eref{Eq10}.
This model highlights besides the effect of the localized Cu$3d$ state
the importance of the delocalized Cu$4s$-component in the charge
response and for the softening of the anomalous OBSM phonon modes. The
importance of the Cu$4s$ orbital for a realistic description of the
electronic structure of the HTSC's has also been pointed out in Refs.
\cite{Andersen95,Pav01}.

In order to demonstrate selectively the renormalization effect
introduced by the Cu$4s$ orbital for the OBSM not only within the
$\Pi$-model approach we have calculated $\Pi_{\kappa\kappa'}$
approximately using for the electronic band structure of the CuO plane
an eleven-band-model (11BM: based on Cu$3d$ and O$2p$ states
\cite{Vielsack90}) and a twelve-band-model (12BM) which takes the
Cu$4s$ state in addition into account.

\begin{figure}%
\centering%
 \includegraphics[width=\linewidth]{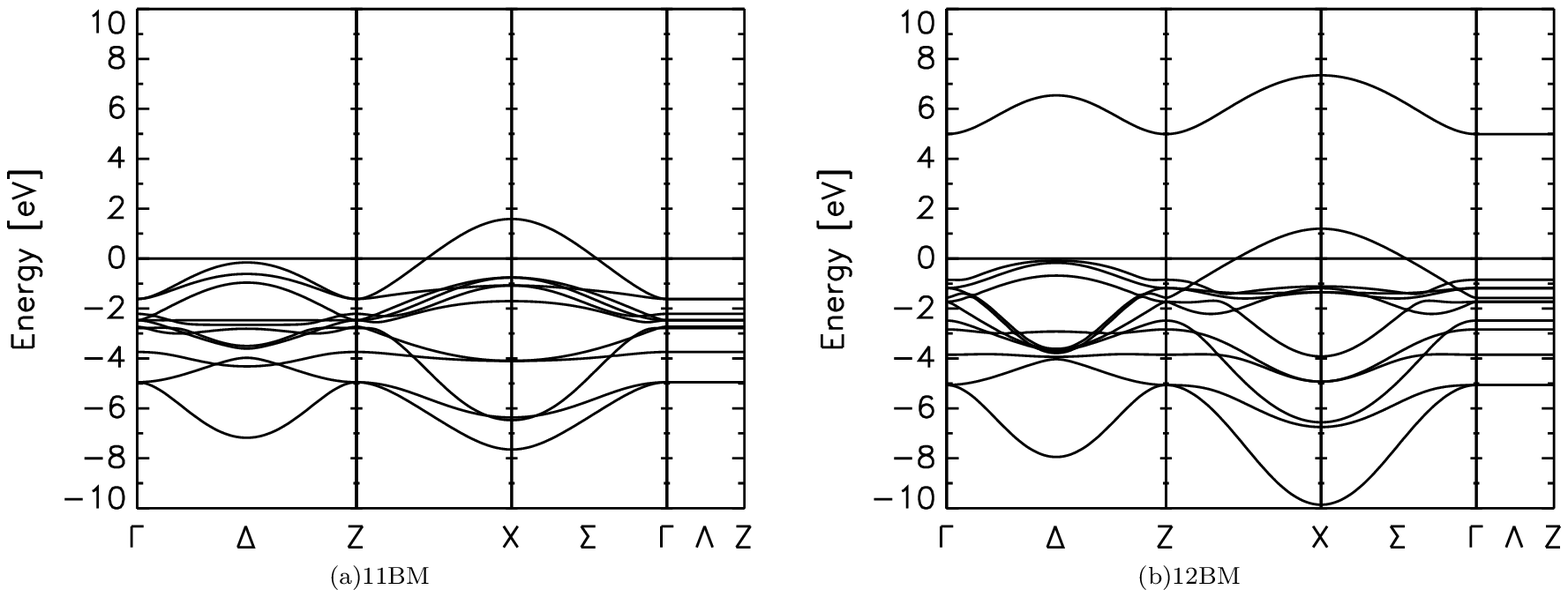}%
 \caption{Electronic band structure of the (a) 11BM  and (b) 12BM.}\label{fig01}%
\end{figure}%

\begin{figure}%
\centering%
 \includegraphics[width=0.3\linewidth]{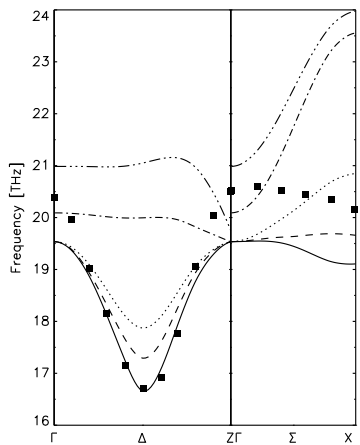}%
 \caption{Calculated OBSM phonon anomalies in La$_{2}$CuO$_{4}$:
 RIM (\dashddot), RIM with anisotropic DF's (\chain), 11BM (\dotted),
 12BM (\broken) and $\Pi$-model (\full). The squares (\fullsquare)
 represent experimental results \cite{Pint98}.}\label{fig02}%
\end{figure}%

In \fref{fig01} the result for the electronic band structure of the
11BM and the 12BM is displayed and in \fref{fig02} the corresponding
results for the OBSM phonon anomalies are shown for the case of
La$_{2}$CuO$_{4}$. The important contribution of the Cu$4s$ degree of
freedom in the charge response for the anomalies is evident by
inspection of the dotted and broken curve in \fref{fig02}. The full
curve relies on a diagonal $\Pi$-model \cite{Falter00} where the matrix
elements of the polarizability matrix $\Pi({\rm Cu}3d)$ = 2.8
eV$^{-1}$, $\Pi({\rm Cu}4s)$ = 0.055 eV$^{-1}$ and $\Pi({\rm
O}_{xy}2p)$ = 0.2 eV$^{-1}$ have been extracted from the full
polarizability matrix, as obtained from a tight-binding analysis of the
electronic band structure for La$_{2}$CuO$_{4}$ (31BM \cite{deWert89})
within the framework  of the local density approximation (LDA) of
density functional theory (DFT). The value for $\Pi({\rm Cu}4s)$,
however, has been slightly increased compared with the 31BM from 0.05
eV$^{-1}$ to 0.055 eV$^{-1}$ in order to optimize the calculated result
for the $\Delta_{1}$ branch in comparison to the experiment.

Using the 11BM or the 12BM, respectively, as generic models for the
electronic structure of the CuO plane also in the calculation of the
phonon dispersion of Nd$_{2}$CuO$_{4}$, HgBa$_{2}$CuO$_{4}$ and
Bi$_{2}$Sr$_{2}$CuO$_{6}$ we find that the softening of the OBSM due to
the Cu$4s$ CF's is of the same size as for La$_{2}$CuO$_{4}$ displayed
in \fref{fig02}.

\begin{figure}%
\centering%
 \includegraphics[width=\linewidth]{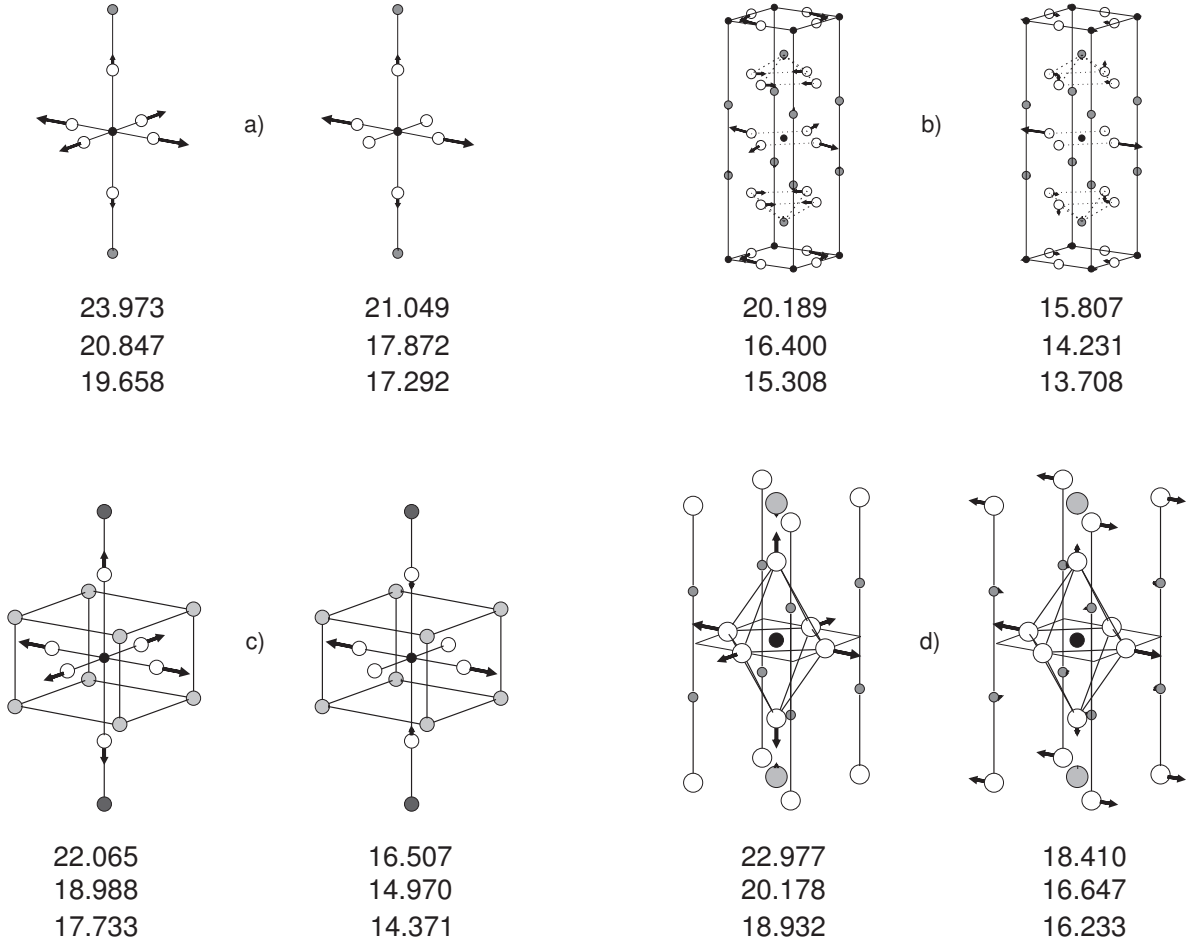}%
 \caption{Displacement patterns of the OBSM for (a) La$_{2}$CuO$_{4}$, (b) Nd$_{2}$CuO$_{4}$,
 (c) HgBa$_{2}$CuO$_{4}$ and (d) Bi$_{2}$Sr$_{2}$CuO$_{6}$. The left pattern
 shows O$_{\rm B}^{X}$ and the right pattern $\Delta_1/2$, respectively. For each displacement pattern
 the phonon frequency is given for the RIM (first row), the 11BM (second row) and the 12BM
 (third row) in units of THz.}\label{fig03}%
\end{figure}%

\begin{table}
\centering
\begin{tabular}{c|ccc|ccc}
             & \multicolumn{3}{c|}{\textbf{La$_2$CuO$_4$}} & \multicolumn{3}{c}{\textbf{Nd$_2$CuO$_4$}} \\
                      & Cu$3d$    & Cu$4s$     & O$_y2p$ & Cu$3d$    & Cu$4s$   & O$_y2p$ \\\hline
O$_{\rm B}^{X}$  & -23.216   &    /       & $-$     & -18.860   & /         & $-$  \\
                      & -19.949 & -13.255  & $-$     & -16.976 & -13.129 & $-$\\
$\Delta_1/2$  & -16.946   &    /       & -6.520  & -14.221   & /         & -5.666  \\
                      & -15.393 & -8.685   & -3.283& -13.284 & -8.473  & -3.152 \\
        & & & & &              \\
             & \multicolumn{3}{c|}{\textbf{HgBa$_{2}$CuO$_{4}$}} & \multicolumn{3}{c}{\textbf{Bi$_{2}$Sr$_{2}$CuO$_{6}$}} \\
                      & Cu$3d$    & Cu$4s$     & O$_y2p$ & Cu$3d$    & Cu$4s$   & O$_y2p$ \\\hline
O$_{\rm B}^{X}$ & -24.231   &    /       & $-$     & -22.078   & /         & $-$  \\
                      & -17.100 & -13.576  & $-$     & -17.081 & -11.575 & $-$\\
$\Delta_1/2$  & -17.528   &    /       & -8.865  & -13.881   & /         & -8.756  \\
                      & -16.276 & -9.228   & -5.904& -11.955 & -6.473  & -4.002 \\
\end{tabular}
\caption{Magnitudes of the calculated charge CF's $\delta\zeta_\kappa$
in units of $10^{-3}$ particles for the breathing and half-breathing
modes. In the first row the values for the 11BM are given, the second
row shows the values for the 12BM, respectively. The symbol / means,
that CF's are not allowed in the model, $-$ means CF's are not excited
for symmetry reasons. The corresponding displacement patterns are shown
in \fref{fig03}.}\label{tab03}
\end{table}

The detailed results for the renormalization of the OBSM are shown in
\fref{fig03} where the displacement patterns of the half-breathing mode
($\Delta_1/2$ anomaly) and the full breathing mode O$_{\rm B}^{X}$ are
given. Moreover, we have listed the corresponding frequencies in THz as
calculated in the RIM, the 11BM and the 12BM. It should be noted, that
due to the simple tetragonal elementary cell, in HgBa$_{2}$CuO$_{4}$
the full breathing mode is located at $M$ and the half-breathing mode
at $X$, but to be consistent with the other three compounds we keep the
denotation $\Delta_1/2$ and O$_{\rm B}^{X}$ also for
HgBa$_{2}$CuO$_{4}$.

A measure for the electron-phonon coupling strength is given by the
mode- and orbital selective charge fluctuations
$\delta\zeta_\kappa(\vc{q}\sigma)$ from equation \eref{Eq11} and the
related renormalization of the corresponding phonon modes. We find a
very strong renormalization of the mode frequencies by the electronic
polarizations processes which nearly exclusively results from the
nonlocally excited CF's $\delta\zeta_{\kappa}$ listed in \tref{tab03}.
The softening for O$_{\rm B}^{X}$ as compared to the RIM is in the
range between 4 THz and 5 THz and for $\Delta_1/2$ between 2 THz and 4
THz. The largest softening for $\Delta_1/2$ occurs in La$_{2}$CuO$_{4}$
partly due to the fact that there is virtually no anticrossing with
other modes of $\Delta_{1}$ symmetry.

The significant renormalization of the OBSM in all the compounds
investigated so far signals a strong nonlocal electron-phonon coupling
in the cuprates. This strong coupling, in particular the effect of the
Cu$4s$ orbital, can also be read off from the calculated magnitude of
the phonon induced charge response given in \tref{tab03} in terms of
the CF's $\delta\zeta_\kappa(\vc{q}\sigma)$ according to equation
\eref{Eq11}.

Our calculations for HgBa$_{2}$CuO$_{4}$ and Bi$_{2}$Sr$_{2}$CuO$_{6}$
in the following, are based on the $\Pi$-model approach because unlike
to La$_{2}$CuO$_{4}$ no reliable tight-binding parametrization of a
first principles band structure is available for these compounds. Even
if a tight-binding parametrization of a typical DFT-LDA based band
structure would exist it would be not sufficient to describe the
dynamics of the $\Lambda_{1}$ modes in the HTSC polarized and
propagating along the $c$-axis. This is, because typical DFT-LDA
calculations are much too isotropic and underestimate the anisotropy of
the real materials by far. For a detailed discussion of this
interesting topic in case of La$_{2}$CuO$_{4}$ and its relevance for
the nonadiabatic charge response and the related phonon-plasmon mixing,
see Refs. \cite{Bauer09a,Bauer09b}. A short comment should be made
concerning the size of the electron-phonon coupling strength of the
adiabatic OBSM with amplitudes within the plane ($\Delta_1/2$, O$_{\rm
B}^{X}$) compared to modes with out of plane amplitudes as the
nonadiabatic apex oxygen bond-stretching mode O$_z^Z$. Like in the
present paper in our calculations for La$_{2}$CuO$_{4}$ \cite{Bauer09a}
an already strong coupling of the adiabatic modes $\Delta_1/2$ and
O$_{\rm B}^{X}$ has been found. However, for the nonadiabatic O$_z^Z$
mode with out of plane amplitudes we obtain in addition a very strong
increase of the coupling as compared to the adiabatic in plane modes.

Phonon-plasmon mixing in the nonadiabatic region around the $c$-axis
provides an extra channel for the formation of cooper pairs. For a
suitable strength of interlayer coupling it becomes possible to achieve
large contributions to the coupling from both, the phonon-like and the
plasmon-like modes, as discussed in \cite{Falter05,Falter02jpc}.
Assuming, that the phonon-plasmon contribution to pairing is
significant this finding could help in the search of new high-$T_c$
materials. This can be accomplished by changing the out of plane
solid-state chemistry in order to tune the interlayer coupling.

\begin{figure}%
\centering%
 \includegraphics[width=\linewidth]{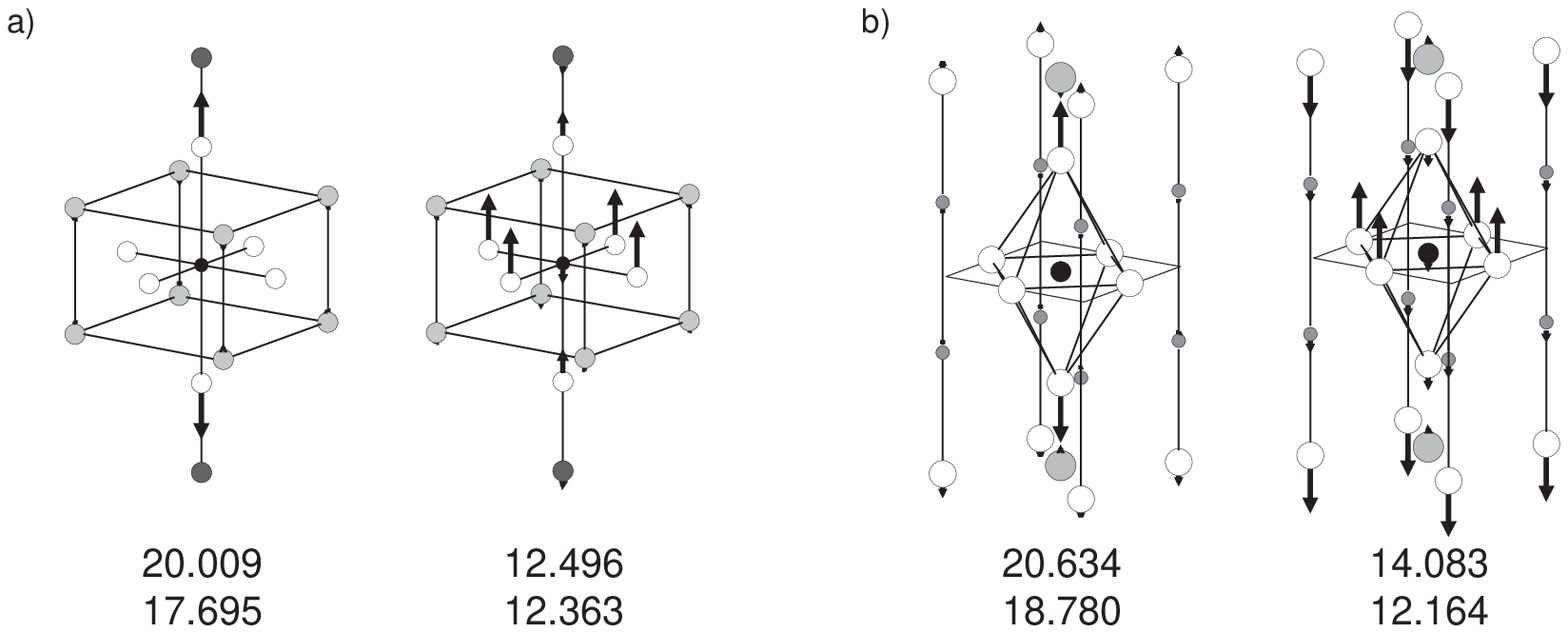}%
 \caption{Displacement patterns of selected optical modes for
 (a) HgBa$_{2}$CuO$_{4}$ and (b) Bi$_{2}$Sr$_{2}$CuO$_{6}$. The left pattern
 shows the Raman active O$^{\Gamma}_{z}$ mode and the right pattern the infrared
 active $A_{2u}$ mode discussed in the text, respectively. For each displacement
 pattern the phonon frequency is given for model $\Pi$-2dim (first row)
 and model $\Pi$-3dim (second row) in units of THz.}\label{fig04}%
\end{figure}%

\begin{table}
\centering
\begin{tabular}{cccccccccc}
HgBa$_{2}$CuO$_{4}$      & $A_{1g}$ & $A_{1g}\,({O_z^\Gamma})$ & $E_{g}$ & $E_{g}$ & $A_{2u}$ & $A_{2u}$ & $A_{2u}$ & $A_{2u}$\\
$\Pi$-2dim              & 5.17     &  20.01   &  2.61   &  4.20   & 3.44     &  4.87    &   12.50  &  17.55\\
$\Pi$-3dim              & 4.63     &  17.70   &  2.61   &  4.20   &3.31    &  4.86    &   12.36  &  17.32\\
Exp. \cite{Krantz94}     & 4.83     &  17.75   &  2.25   &  5.04\\
Exp. \cite{Singley09}    &          &          &         &         & 2.58 &4.50 & 10.67 & 18.80\\
\\
Bi$_{2}$Sr$_{2}$CuO$_{6}$& $A_{1g}$ & $A_{1g}$ & $A_{1g}$ & $A_{1g}\,({O_z^\Gamma})$ & $A_{2u}$ & $A_{2u}$ & $A_{2u}$ & $A_{2u}$ & $A_{2u}$ \\
$\Pi$-2dim              &  4.18    &  7.02    &  14.06   &  20.63   &  2.99    &  5.09    &  9.25    &   14.08  &  18.67 \\
$\Pi$-3dim              &  4.16    &  5.89    &  13.59   &  18.78   &  2.79    &  5.08    &  9.17    &   12.16  &  18.10  \\
Exp. \cite{Osada97}      &  2.07    &  3.57    &  13.61   &  18.71\\
Exp. \cite{Liu92}        & 3.60     &  6.00    &  13.76   &  18.74\\
Exp. \cite{Kovaleva04}   &          &          &          & & 3.27   & 4.95     &  9.02    &  11.54   &  17.57\\
\end{tabular}
\caption{Frequencies of the calculated and experimental measured Raman-
and infrared-active modes in units of THz.}\label{tab04}
\end{table}

Within our $\Pi$-model approach  we can model the charge response for
ions out of the CuO plane by adjusting the corresponding polarizability
matrix elements of these ions in order to obtain a reasonable
description of the frequency of certain phonon modes which are most
sensitive to the charge response of the most important out-of-plane
ions. So, as a first approximation we take the $\Pi$-model for the CuO
plane as introduced for La$_{2}$CuO$_{4}$ above also for
HgBa$_{2}$CuO$_{4}$ and Bi$_{2}$Sr$_{2}$CuO$_{6}$. From \tref{tab04} a
comparison of the calculated frequencies of the Raman modes with the
experiment shows that the largest difference between theory and
experiment appears for the Raman mode O$^{\Gamma}_{z}$ with the highest
frequency. Here the apex oxygen ion vibrates along the $c$-axis against
the Hg or Bi ion, respectively, which are located just above the apex
oxygen, see \fref{fig04}. As already mentioned in section 3.1 there is
a relatively short distance between O$_{z}$ and Hg or Bi, respectively,
and consequently metallic CF's can be expected which should renormalize
the corresponding modes.

In order to investigate this effect we extend the $\Pi$-model for the
CuO plane ($\Pi$-2dim) by allowing for metallic charge fluctuations on
the O$_{z}$, Hg and Bi ions ($\Pi$-3dim). The corresponding matrix
elements $\Pi({\rm Hg}6s)$, $\Pi({\rm O}_{z}2p)$, for
HgBa$_{2}$CuO$_{4}$ and $\Pi({\rm Bi}6p)$, $\Pi({\rm O}_{z}2p)$ for
Bi$_{2}$Sr$_{2}$CuO$_{6}$ are determined in such a way that the
calculated and measured frequencies for O$^{\Gamma}_{z}$ are very
close, see \tref{tab04} and \fref{fig04}.

Moreover, we extract from the results in \tref{tab04} that also the
other calculated Raman modes are improved by allowing CF's at these
ions out of the CuO plane. In case of Bi$_{2}$Sr$_{2}$CuO$_{6}$ it is
remarkable that also an infrared active $A_{2u}$ mode is softened by
nearly 2 THz if the out-of-plane CF's are in addition admitted. As can
be expected, in this mode the O$_z$ and the Bi ion are vibrating
against each other, compare with \fref{fig04}. It should be noted that
in calculations of the infrared and Raman active modes for
La$_{2}$CuO$_{4}$ and Nd$_{2}$CuO$_{4}$ with CF's allowed in the
$\Pi$-model on O$_{z}$, La or Nd, respectively, these CF only do have a
minor influence on the phonon frequencies.

\begin{figure}%
\centering%
 \includegraphics[height=\linewidth,angle=90]{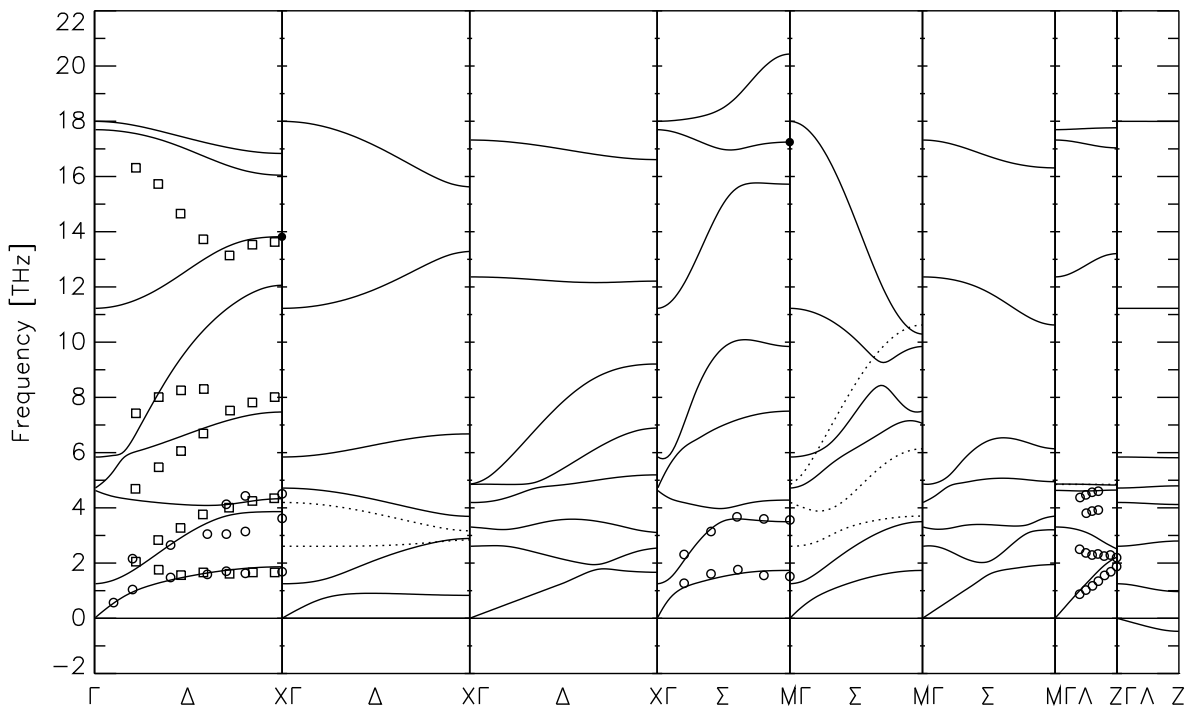}%
 \caption{Calculated phonon dispersion of HgBa$_{2}$CuO$_{4}$ for model $\Pi$-3dim in
 the main symmetry directions $\Delta\sim(1,0,0)$, $\Sigma\sim(1,1,0)$ and
 $\Lambda\sim(0,0,1)$. The symbols $\circ$ and $\Box$ represent experimental
 results from \cite{d'Ast03} and \cite{Uchi04}, respectively. The circle $\fullcircle$
 marks two breathing modes at $M$ and the half-breathing mode at $X$.
 The arrangement of the panels from left to right according to the different
 irreducible representations is as follows:$|\Delta_{1}|\Delta_{2}$ (\dotted),
 $\Delta_{4}$ (\full)$|\Delta_{3}|\Sigma_{1}|\Sigma_{2}$ (\dotted), $\Sigma_{4}$ (\full)
 $|\Sigma_{3}|\Lambda_{1}$ (\full), $\Lambda_{2}$ (\dotted)$|\Lambda_{3}|$.
 Imaginary frequencies are represented as negative numbers.}\label{fig05}%
\end{figure}%

\begin{figure}%
\centering%
 \includegraphics[height=\linewidth,angle=90]{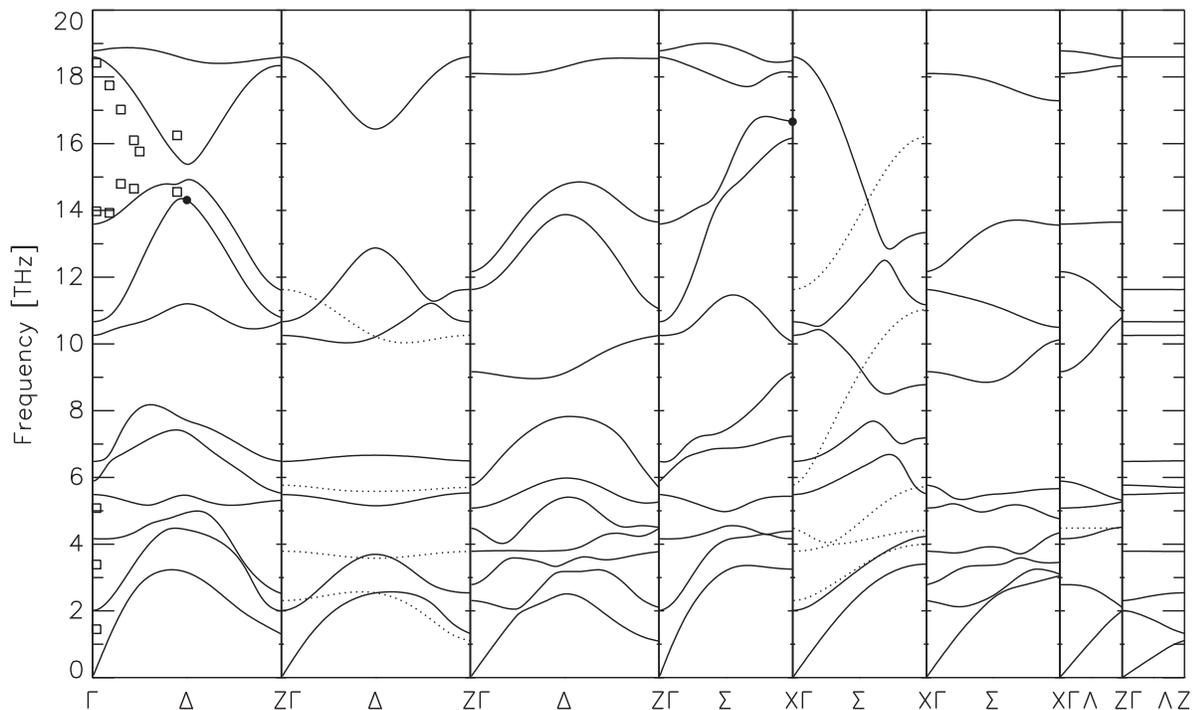}%
 \caption{Calculated phonon dispersion of Bi$_{2}$Sr$_{2}$CuO$_{6}$ for model $\Pi$-3dim in
 the main symmetry directions $\Delta\sim(1,0,0)$, $\Sigma\sim(1,1,0)$ and
 $\Lambda\sim(0,0,1)$. The symbols $\Box$ represent experimental
 results from \cite{Graf08}. The circle $\fullcircle$
 marks the breathing modes at $X$ and the half-breathing modes at $\Delta/2$. For the arrangement of the
 panels see \fref{fig05}.}\label{fig06}%
\end{figure}%

In \fref{fig05} and \ref{fig06} the calculated phonon dispersion of
HgBa$_{2}$CuO$_{4}$ and Bi$_{2}$Sr$_{2}$CuO$_{6}$ is displayed in the
main-symmetry direction $\Delta \sim (1,0,0)$, $\Sigma \sim (1,1,0)$
and $\Lambda \sim (0,0,1)$. In addition, an interpretation of the
IXS-phonon spectra for HgBa$_{2}$CuO$_{4}$ \cite{Uchi04,d'Ast03} and
for Bi$_{2}$Sr$_{2}$CuO$_{6}$ \cite{Graf08} is added to the figures in
form of open squares and circles.

In case of HgBa$_{2}$CuO$_{4}$ a minor deficit is visible in form of
one slightly unstable acoustic $\Lambda_{3}$ mode which is related to
lattice vibrations propagating along the $c$-axis where the O$_{z}$ and
Ba ions are vibrating in phase parallel to the CuO plane. This
instability is already present in the RIM and most likely is a result
of the more open structure of HgBa$_{2}$CuO$_{4}$ and thus of the the
growing importance of non-central forces.

\begin{figure}%
\centering%
 \includegraphics[width=0.9\linewidth]{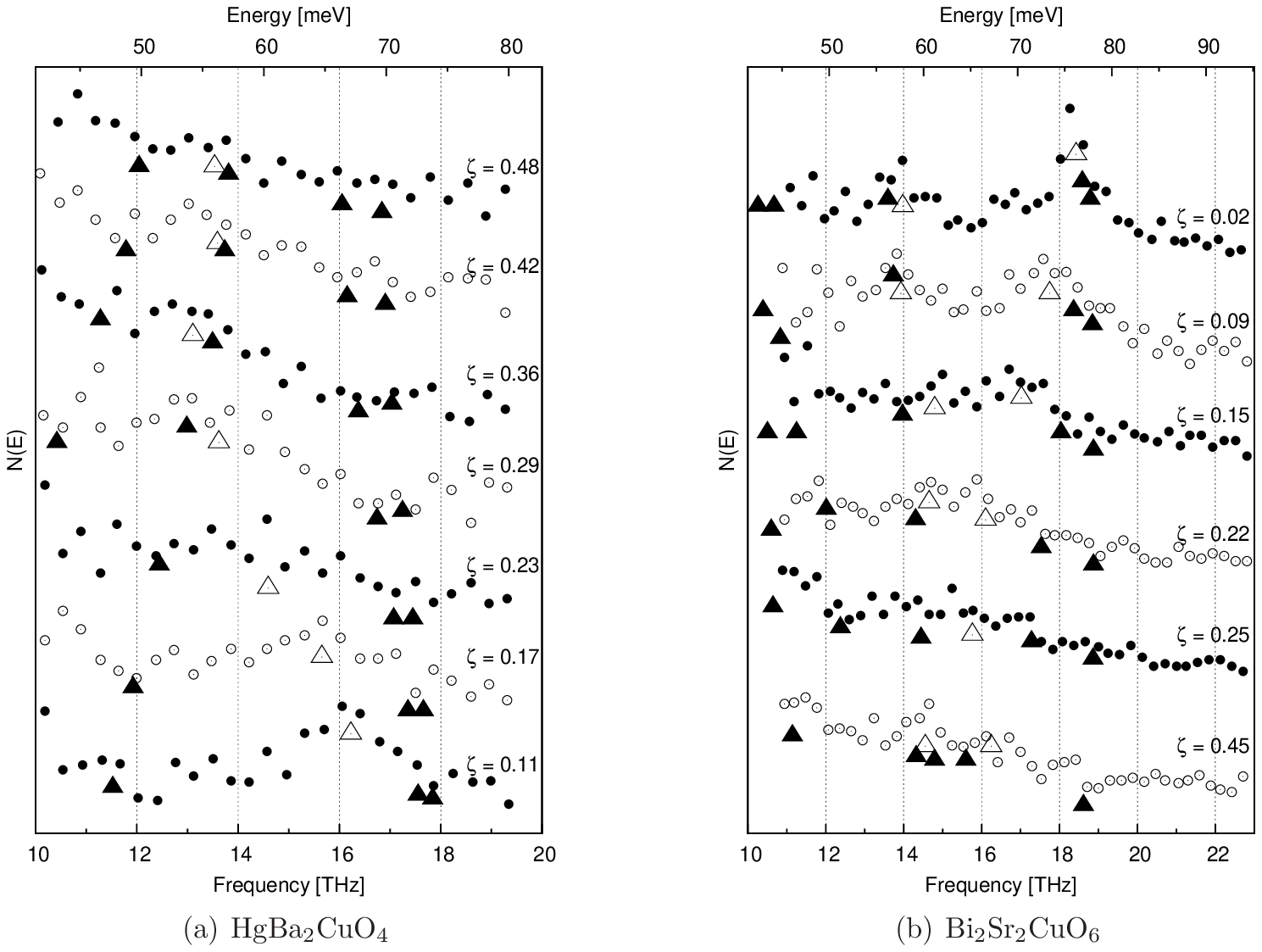}%
 \caption{Reconstructed experimental x-ray spectra $N(E)$ of (a) HgBa$_{2}$CuO$_{4}$
 for $\vc{q}=\frac{\pi}{a}(\zeta,0,0)$ \cite{Uchi04}
 and (b) Bi$_{2}$Sr$_{2}$CuO$_{6}$ for $\vc{q}=\frac{2\pi}{a}(\zeta,0,0)$ \cite{Graf08}
 in arbitrary units, respectively.
 The experimental spectra for the different $\zeta$ are represented by the open and closed circles
 (\opencircle,\fullcircle). The according interpretations of the
 phonon frequencies fitted by a Voigt fit \cite{Uchi04,Graf08} are represented
 by open triangles (\opentriangle). Our calculated phonon
 frequencies are represented by closed triangles (\fulltriangle).}\label{fig07}%
\end{figure}%

The interpretation of the IXS-spectra in Refs. \cite{Uchi04,d'Ast03}
and \cite{Graf08}, respectively, are based on a Voigt fit. The
positions of the peaks resulting from this fit are displayed in the
IXS-spectra shown in \fref{fig07}(a) for HgBa$_{2}$CuO$_{4}$ and
\ref{fig07}(b) for Bi$_{2}$Sr$_{2}$CuO$_{6}$ as open triangles. In
addition, we have plotted in this figure the calculated frequencies of
the $\Delta_{1}$-modes in the energy range considered, including the
half-breathing mode, at the corresponding wavevectors (full triangles).

At first one notices that for HgBa$_{2}$CuO$_{4}$ only one phonon peak
has been fitted in the energy range shown in \fref{fig07}(a). On the
other hand, we find in this range of energy a phonon spectrum which is
much more complex comprising up to four $\Delta_{1}$ modes, see
\fref{fig05}. In the fit for Bi$_{2}$Sr$_{2}$CuO$_{6}$ \cite{Graf08}
only two phonon peaks have been fitted in the range of energy shown in
\fref{fig07}(b), while according to our calculations in \fref{fig06}
there are up to five $\Delta_{1}$ modes in this range which can
interact with each other and constitute a highly nontrivial
anticrossing scenarium. Quite general we can say that the IXS-spectra
reveal for small $\zeta$ values $(\vc{q} = (\zeta,0,0))$ in
HgBa$_{2}$CuO$_{4}$ one and in Bi$_{2}$Sr$_{2}$CuO$_{6}$ two marked
peaks. However, the strength of the peaks strongly decreases for larger
$\zeta$ values such that at $\zeta \approx 0.5$ the peaks are very weak
and an unique interpretation seems not possible. Moreover, the dips in
the phonon dispersion curves as extracted from the interpretation of
the IXS-spectra at $\zeta = 0.36$ for HgBa$_{2}$CuO$_{4}$ \cite{Uchi04}
and at $\zeta = 0.25$ for Bi$_{2}$Sr$_{2}$CuO$_{6}$ \cite{Graf08} are
not found in our calculations. However, as a result of the complex
anticrossing scenario we obtain a pronounced dip at $\zeta = 0.5$ in
Bi$_{2}$Sr$_{2}$CuO$_{6}$ and two half-breathing modes below the dip
with somewhat smaller frequencies, see \fref{fig06}. The higher one has
a displacement pattern as shown in \fref{fig03} (d) and in the lower
one the apex oxygen and the oxygens in the ionic layer vibrate with
opposite phases compared to \fref{fig03} (d). This holds also true for
the $X$-point, where two breathing modes occur, see \fref{fig06}. The
higher one has the displacement pattern as shown in \fref{fig03} (d),
and in the lower one the apex oxygen and the Bismuth ion vibrate with
opposite phases compared to \fref{fig03} (d).

In case of HgBa$_{2}$CuO$_{4}$ we do not find a pronounced dip for the
$\Delta_{1}$ modes in the relevant range of energy. We obtain as a
result of four interacting $\Delta_{1}$ modes two branches with the
highest frequencies showing a downward dispersion from $\Gamma$ to $X$
while the anomalous half-breathing mode is found at a lower energy and
turns out to be the end point of the $\Delta_{1}$ branch with the third
highest frequency. Similar as in the case of Bi$_{2}$Sr$_{2}$CuO$_{6}$
we find in HgBa$_{2}$CuO$_{4}$ two breathing modes at the $M$ point.
The lower one has a displacement pattern where the apex oxygen vibrates
in opposite direction as compared to the higher frequency breathing
mode displayed in figure \fref{fig03} (c).

So, we guess from our calculations for HgBa$_{2}$CuO$_{4}$ and
Bi$_{2}$Sr$_{2}$CuO$_{6}$ that the dip in the interpretation of the
IXS-spectra may be an artefact of the complex anticrossing behaviour of
several $\Delta_{1}$ modes in the frequency range under consideration.
However, we admit concerning our results, that our microscopic model
for the electronic polarizability applied to the calculation of the
phonon dynamics of HgBa$_{2}$CuO$_{4}$ and Bi$_{2}$Sr$_{2}$CuO$_{6}$ is
not fully ab initio.

The reliability of our calculated phonon dispersion may be judged from
the conformance of our results with the phonon branches in
HgBa$_{2}$CuO$_{4}$ at lower frequencies measured so far and displayed
in \fref{fig05} for the model $\Pi$-3dim.

The experimental and calculated results for the Raman and infrared
active modes \cite{Krantz94,Osada97,Liu92,Singley09,Kovaleva04} are
listed in \tref{tab04} for both compounds and a good agreement is found
for the measured and the calculated data in model $\Pi$-3dim. This fact
together with the good results in model $\Pi$-3dim for the dispersion
underlines the importance of the electronic polarization processes of
certain ions in the ionic layers and in general of the physics of the
third dimension perpendicular to the CuO plane.

\begin{figure}%
\centering%
 \includegraphics[height=\linewidth,angle=90]{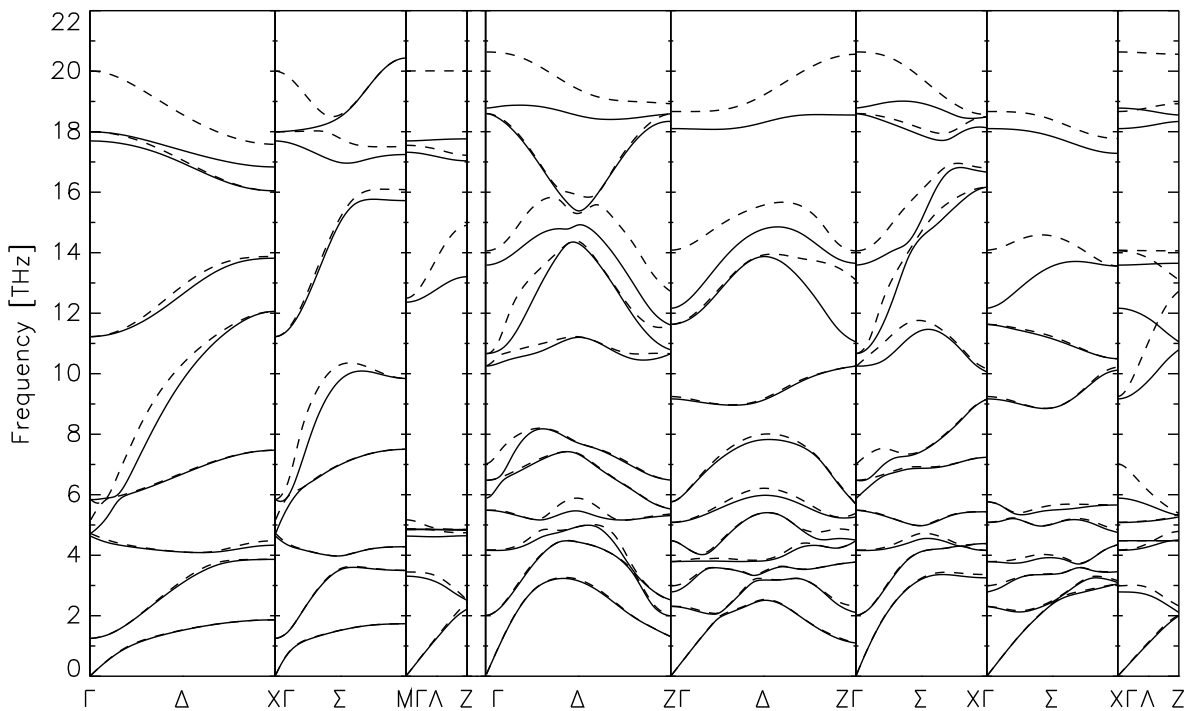}%
 \caption{Comparison of model $\Pi$-2dim (\dashed) and model
 $\Pi$-3dim (\full). The three leftmost panels belong to
 HgBa$_{2}$CuO$_{4}$ and the five rightmost panels belong to
 Bi$_{2}$Sr$_{2}$CuO$_{6}$. Only irreducible representations with a
 different phonon dispersion in model $\Pi$-2dim and model $\Pi$-3dim are shown.
 The arrangement of the panels from
  left to right according to the different
 irreducible representations is as follows:
 $|\Delta_{1}|\Sigma_{1}|\Lambda_{1}||\Delta_{1}|
 \Delta_{3}|\Sigma_{1}|\Sigma_{3}|\Lambda_{1}|$.}\label{fig08}%
\end{figure}%

The effect introduced by the CF's on the O$_{z}$, Hg and Bi ions
outside of the CuO plane on the phonon dispersion by itself is
demonstrated in \fref{fig08} for both compounds by comparing the
results of model $\Pi$-2dim (broken lines) with those of model
$\Pi$-3dim (full lines). Only the branches are shown where large
differences appear. The most significant renormalization of the
dispersion emerges for the modes with the highest frequencies.

\begin{figure}%
\centering%
 \includegraphics[width=\linewidth]{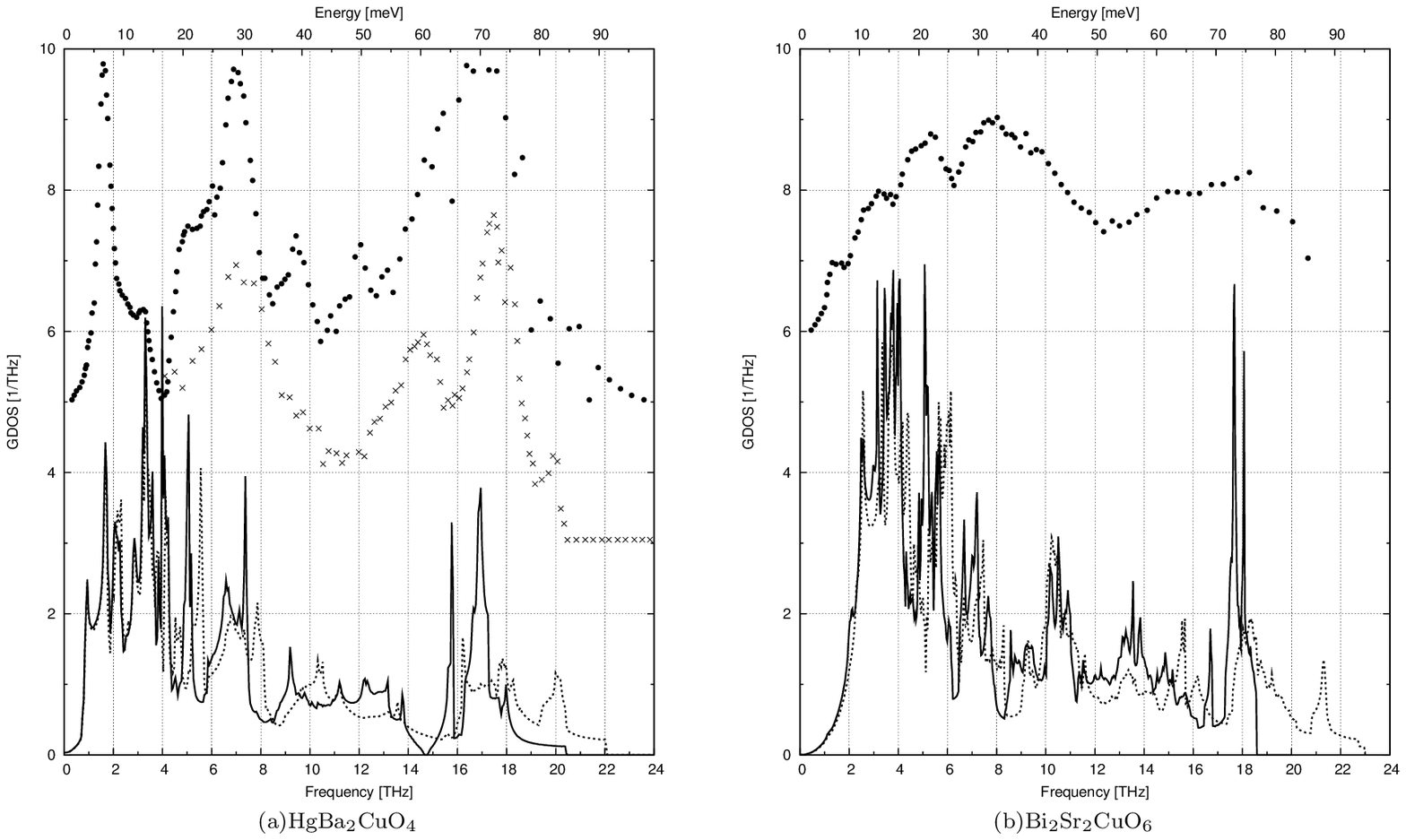}%
 \caption{Calculated phonon density of states (DOS) for (a) Hg$_{2}$BaCuO$_{4}$
  and (b) Bi$_{2}$Sr$_{2}$CuO$_{6}$ for model $\Pi$-3dim (\full).
  For comparison, the DOS of the RIM is shown as dotted line (\dotted). The symbols
  $\times$ and $\bullet$ in (a) show the experimental data \cite{Renker96} at 30 and 300K, respectively,
  the symbols $\bullet$ in (b) represent the experimental data \cite{Parshin96} at 300K. All
  experimental data are in arbitrary units and shifted for better viewing.}\label{fig09}%
\end{figure}%

Finally, in \fref{fig09}, the calculated phonon density of states (DOS)
is shown for HgBa$_{2}$CuO$_{4}$ and Bi$_{2}$Sr$_{2}$CuO$_{6}$,
respectively, and is compared with the experimental findings for both
compounds. The theoretical results have been obtained with model
$\Pi$-3dim. They compare quite well with the experimental results for
both, the width of the spectra and their peak structure. Of course the
assignment of the peaks is difficult in detail because the experimental
data are broadened.

A calculation of the partial phonon density of states, not shown in
this work, demonstrates that the DOS in the lower frequency range is
dominated by Cu-, Ba-, Hg-vibrations and by Cu-, Bi-, Sr-vibrations,
respectively. On the other hand, the high-frequency part of the
spectrum is nearly exclusively governed by the oxygen modes.

From \fref{fig09} (a) and (b) also the difference of the DOS in the
model $\Pi$-3dim and the RIM generated by the electronic polarization
processes can be extracted. From the calculated results we find a
significant decrease of the width of the spectra together with a
characteristic redistribution of spectral weight particularly in the
high-frequency part. The latter is dominated by strongly coupling
high-frequency oxgen vibrations living in this part of the spectra and
suffering a corresponding softening compared to the RIM. The pronounced
peaks which are generated by this softening also mean that there is
considerable phase space for this strongly coupling modes.

\section{Summary and conclusions} \label{SecFour}
We have shown that for a reliable description of phonon dynamics and
charge response in HgBa$_{2}$CuO$_{4}$ and Bi$_{2}$Sr$_{2}$CuO$_{6}$ a
broad set of orbital degrees of freedom (Cu$3d$, Cu$4s$, O$2p$) must be
taken into account to model the electronic structure of the CuO plane.
Moreover, for an ample representation of the real three-dimensional
compounds electronic degrees of freedom beyond the strictly
two-dimensional electronic structure must be considered.

We found that in the link between the CuO plane and the ionic layers of
the cuprates electronic polarization processes of CF type are essential
at certain ions. In HgBa$_{2}$CuO$_{4}$ O$_{z}2p$ and Hg$6s$ CF's and
in Bi$_{2}$Sr$_{2}$CuO$_{6}$ O$_{z}2p$ and Bi$6p$ CF's, respectively,
are significant processes. CF's on these ions are promoted by the short
distance of the O$_{z}$-Hg and O$_{z}$-Bi bond. Allowance of CF's on
these ions outside of the CuO plane (model $\Pi$-3dim) leads to a much
better depiction of the phonon dispersion curves, the Raman active and
the infrared active modes, in particular for those where the apex
oxygen vibrates along its bond with the Hg or the Bi ion, respectively.

As far as the electronic degrees of freedom in the CuO plane are
concerned we have pointed out that besides the localized Cu$3d$ state
and the O$2p$ state also the more delocalized Cu$4s$ state is needed in
the charge response for an adequate representation of the large
softening of the high-frequency oxygen bond-stretching modes and the
strong coupling to the electrons. Such an effect also has been found in
our previous calculations for La$_{2}$CuO$_{4}$ and Nd$_{2}$CuO$_{4}$.
The strong nonlocal coupling and the related strong renormalization of
these modes seems to be generic in the cuprates.

Altogether, our calculated results for model $\Pi$-3dim are partly in
good agreement with the phonon dispersion measured so far in
HgBa$_{2}$CuO$_{4}$ and Bi$_{2}$Sr$_{2}$CuO$_{6}$ and also with the
Raman active and infrared active modes.

However, we have pointed out that an interpretation of the IXS
measurements of HgBa$_{2}$CuO$_{4}$ and Bi$_{2}$Sr$_{2}$CuO$_{6}$ in
the high-frequency part of the spectrum and the assignment of dips in
the corresponding phonon dispersion curves cannot be supported by our
calculations. This is, because our investigations exhibit that in the
frequency range under study we have multiple interacting phonon modes
of $\Delta_{1}$ symmetry (four in HgBa$_{2}$CuO$_{4}$ and five in
Bi$_{2}$Sr$_{2}$CuO$_{6}$) which create a complex anticrossing scenario
that possibly cannot be studied correctly within a fit using just one
or two modes.

Finally, we have calculated the DOS of both compounds in satisfactory
agreement with experiment. We found a significant redistribution of
spectral weight due to the electronic polarization processes in
particular in the high-frequency part of the spectrum. This energy
range is dominated by the strongly coupling oxygen modes signaling a
considerable amount of phase space for strong electron phonon coupling.

%%%%%%%%%%%%%%%%%%%%%%%%%%%%%%%%%%%%%%%%%%%%%%%%%%%%%%%%%%%%%%%%%%%%%
%%%%%%%%%
%%%%%%%%%                                        References, 3rd November 2009
%%%%%%%%%
%%%%%%%%%%%%%%%%%%%%%%%%%%%%%%%%%%%%%%%%%%%%%%%%%%%%%%%%%%%%%%%%%%%%%

\end{document}